\documentclass[11pt]{article}
\usepackage{cite}
\def\be#1{\begin{equation}\label{#1}}

          \def\ee{\end{equation}}

\title{Revisiting Gribov's Copies Inside The Horizon}
\author{R. R. Landim$^{a}$\footnote{email: renan@fisica.ufc.br} ,
V. E. R. Lemes$^{b}$\footnote{email: vitor@dft.if.uerj.br},
O. S. Ventura$^{c}$\footnote{email: ozemar.ventura@cefet-rj.br} ,
L. C. Q. Vilar$^{b}$\footnote{email: lcqvilar@gmail.com}  \\
\small \em $^a$Departamento de F\'{\i}sica, Universidade Federal do Cear\'{a}\\
\small \em Caixa Postal 6030, Campus do Pici, 60455-760, Fortaleza - Cear\'{a},
Brazil\\
\small \em $^b$Instituto de F\'\i sica, Universidade do Estado do Rio de
Janeiro,\\
\small \em Rua S\~{a}o Francisco Xavier 524, Maracan\~{a}, Rio de Janeiro - RJ,
20550-013, Brazil\\
\small \em $^c$Centro Federal de Educa\c{c}\~ao Tecnol\'ogica do Rio de
Janeiro\\
\small\em Av.Maracan\~a 249, 20271-110, Rio de Janeiro - RJ, Brazil}

\begin{document}
\maketitle
\begin{abstract}
In this work, we recover the problem of legitimate topologically trivial Gribov copies inside the Gribov horizon. We avoid the reducibility problem which hampered the standard construction of van Baal, and then we are able to build a valid example with spherical symmetry. We also apply the same technique in the presence of a background of a Polyakov instanton in a Euclidian 3D spacetime, in order to study the effect of a non trivial environment in the generation of multiple copies inside the horizon.
\end{abstract}

\section{Introduction}

Soon after Gribov's seminal work \cite{Gribov:1977wm}, the alleged absence of gauge copies 
in his first region began to be questioned. In \cite{sciuto} Sciuto already mentioned the possible presence 
of gauge copies inside the first region.  Afterwards, by using the formulation of the gauge fixing 
procedure as an action principle, Semenov-Tyan-Shanskii and Franke \cite{semenov:1982} were able 
to argue consistently for the presence of these copies for the first time. The idea was to give an appropriate
topography $I_S$ to each leaf defined by the gauge orbit of a given gauge configuration in the 
foliation of the configuration space. In the case of a Landau gauge fixing,  for example, such 
topography would be given by the Hilbert norm of the gauge field in each point of the gauge orbit

\begin{equation}
I_S={Tr\int_{M} {\tilde{A}_i}^2}	,
\label{Adois}
\end{equation}
where $M$ is the space-time manifold, and $\tilde{A}_i$ represents the gauge connection on the 
orbit generated by the gauge transformations $S$ of a gauge configuration ${A}_i$

\begin{eqnarray}
\tilde{A}_i = S^{\dagger}\partial_i S+S^{\dagger}A_iS	. 
\label{atilde}
\end{eqnarray}
The vanishing of the first variation of this norm functional gives Landau gauge fixing condition. 
So, gauge connections satisfying such condition are in fact extrema of (\ref{Adois}). The next term in the
expansion, i.e. the Hessian of (\ref{Adois}), gives the Faddeev-Popov operator. Then, the restriction implemented by 
Gribov on the configuration space to his first region, $C_{0}$,  of the connections with only positive eigenvalues 
of this operator, means that we are choosing the configurations which are local minima of (\ref{Adois}). The Gribov 
horizon, $l_{1}$, is the boundary of this region, where the lowest eigenvalue of the Faddeev-Popov operator 
vanishes. Then, in the next regions $C_{n}$ we will find $n$ negative eigenvalues (with all others strictly positive), 
each region separated from the next by the Gribov boundaries, $l_{n+1}$, where new zero eigenvalues reappears (details 
of Gribov's construction can be found in the review \cite{silvioreview}). Basically, in \cite{semenov:1982}, the authors argued that there would be no reason for not expecting multiple local minima of (\ref{Adois}) in the same orbit. Consequently, one should expect Gribov copies in $C_{0}$.

It is important here to mention the fundamental result of Dell'Antonio and Zwanziger 
proving that every gauge orbit has at least one copy in $C_{0}$ \cite{delant}. This showed that 
the restriction of the configuration space
to this region would not cut out any physical configuration from the action functionals.

The next development made in this discussion was done by van Baal \cite{1992}.
His work begins by interpreting the question 
of Gribov copies in this variational form as a problem in Morse theory (see also \cite{labast} for a 
previous use of Morse theory on topological quantum field 
theories and its relation to the Gribov ambiguity). In a few words, Morse theory searches for a 
characterization of topological invariants of any given manifold by the study of the critical points 
of functions defined on it \cite{milnor,matsumoto}. In this interpretation, 
the basis manifold would be the gauge orbit, 
and the function defined on it would be the Hilbert norm (\ref{Adois}) 
(actually, a proper formalization of these ideas would require the use of a generalization of the Morse theory for 
functions with degenerate critical points 
\cite{attyah}). 
So, given a gauge field configuration and its gauge orbit, and once the topography  in (\ref{Adois}) is specified for it, one 
can use the results of Morse theory to gain a new perspective on Gribov's problem (at least in a local way). The best 
known result of Morse theory is the expression of the Euler's characteristic $\chi$ of a manifold in terms of the 
critical points of a Morse function defined on it,
\begin{eqnarray}
\chi=\sum_{\mu}(-1)^{\mu}N_{\mu}	,
\label{Eulercha}
\end{eqnarray}
where $N_{\mu}$ is the number of critical points with Morse index $\mu$. The Morse index is the number of negative 
eigenvalues of the Hessian of the Morse function. Applied to (\ref{Adois}), by the definitions of the Gribov's regions,
we see that $\mu$ in this case is exactly the label $n$ of $C_n$, i.e., a region $C_n$ can also be characterized by
stating that it collects all critical non degenerate gauge configurations with Morse index ${\mu}=n$ of the function (\ref{Adois}) in all gauge 
orbits. In \cite{1992}, van Baal uses Morse's result locally. Being a topological invariant, Euler's characteristic  
should not change in continuous deformations of a gauge orbit. So, if in a small neighborhood of a gauge configuration
in $C_0$, i.e. an isolated local minimum of (\ref{Adois}), we proceed with a continuous deformation in such a way to generate a 
local maximum, i.e. a configuration in $C_1$, we will be ultimately requested to preserve $\chi$. As any isolated 
configuration in $C_0$ contributes with $1$ to $\chi$, and as any configuration in $C_1$ 
contributes with $-1$, the way to preserve $\chi$ is by generating two local minima in the same gauge orbit around
the maximum in the deformation process. We observe that with this deformation we are in fact moving in the 
configuration space, transversely in relation to the gauge orbit where the minimum configuration can be found,
and towards the gauge orbit where the maximum belongs to (and certainly crossing a gauge orbit with a configuration at $l_{1}$ in this process). Then, 
if we are able to successfully implement such deformation, the conclusion is that we will find two gauge copies 
inside the first region $C_0$. 

After this reasoning, van Baal started to look for an example of these copies. One important point that
should be mentioned is that he was mainly interested in gauge copies with vanishing winding number. In fact, 
several criticisms to Gribov's work had been previously based on the idea that Gribov copies were an exclusive 
feature of improper large gauge transformations, carrying a non vanishing winding number (see for instance 
\cite{ramond}). 
Following this idea, all the construction on the restriction of the configuration space 
could be irrelevant, since such ambiguities of the gauge potential could be allowed so as to accommodate 
field configurations with non null Pontryagin index \cite{jackiw1978,benguria1977}. 
In order to prove that copies could be found even among 
homotopically trivial gauge configurations, Henyey constructed an explicit example 
\cite{henyey}. The 
strategy designed by van Baal was then to start with Henyey's 
example and show that it fit the deformation picture that he apprehended from Morse. The conclusion then 
was that two copies with vanishing winding number would be generated inside Gribov's horizon \cite{1992}. 

This achievement apparently destroyed the hope that the first region would be free of topologically trivial copies. Although the 
argument based on Morse theory strongly indicates their presence, actually the example built from Henyey's 
configuration was unfortunate. Let us show why this is so. Let us write two connections obtained from different 
gauge transformations of the same initial configuration ${A}_i$, from different SU(2) gauge transformations

\begin{eqnarray}
\tilde{A}_i = S^{\dagger}\partial_i S+S^{\dagger}A_iS  \hspace{1cm}	, \hspace{1cm}  \tilde{A}_i^{'} = U^{\dagger}\partial_i U+U^{\dagger}A_iU .
\label{inicialAtilde} 
\end{eqnarray}
However, if ${A}_i$ is a reducible configuration, i.e., if there is a generator $T$ in the gauge group such that in the SU(2) gauge group
\begin{eqnarray}
[T,A_i] = 0,
\label{(5)}
\end{eqnarray}
and if at the same time the transformations $S$ and $U$ are related by $T$,
\begin{eqnarray}
U={T^\dagger}ST ,
\label{(6)}
\end{eqnarray}
then it is immediate to see that $\tilde{A}_i$ and $\tilde{A}_i^{'}$ are in fact related by a global gauge transformation, 
\begin{eqnarray}
\tilde{A}_i^{'}=T^\dagger\tilde A_i T .
\label{(7)}
\end{eqnarray}
The point is that Henyey's configuration is reducible and the gauge transformations generating the copies 
inside  the first horizon satisfy (\ref{(6)}). Then, it just happens that the copies found in \cite{1992} are global gauge
copies. This way, although meeting the standards of Morse theory, they cannot be considered as legitimate Gribov copies. 
This final result was actually noticed by van Baal. In \cite{9207029}, after sketching this argument, he 
reinforces his belief in the existence of an initial irreducible gauge configuration, instead of Henyey's, able 
to induce the kind of bifurcation suggested by Morse theory. In subsequent works, copies inside the horizon for 
sphaleron configurations were produced for gauge theory on a three sphere space-time  \cite{9310005,9511119,9711070}
with the cost of abandoning then the homotopical 
triviality demand for the gauge configurations. 

We should also mention that presently the existence of Gribov copies inside the first region has also strong evidence coming from lattice simulations. These studies started even before van Baal's work on Morse theory \cite{1991,Parrinello,Vladikas,Sinclair}, but even nowadays it is systematically pointed out that some of these copies are just lattice artifacts \cite{attilio}. It is unlikely that all lattice copies would disappear in the continuum theory, although no formal proof that this is the case has been shown up to now (specially for the zero winding number configurations inside the first region). But it must be highlighted the recent result that numerical simulations indicate that the lattice theory shows different deep infrared behaviours for expectation values calculated over configurations in the Fundamental Modular Region, free of Gribov copies, from those obtained in the first Gribov region \cite{bornyakov}, showing once more the relevance of the analytical identification of Gribov copies inside the first region in the continuum.

Anyway, the fake copies shown in \cite{1992} are very useful to reveal that the argument on Morse theory 
is not enough to ensure the presence of copies with vanishing winding number inside the Gribov's horizon. 
The question of rigid gauge copies, or the asymptotic demands on gauge fields and transformations, which 
we shall address in the following, are issues among the physical restrictions which need to be imposed on the 
results coming from the basic appliance of Morse theory.
Then, this is the first goal of our work, to provide a simple example of such gauge copies  inside $C_0$ for 
an Euclidian $R^3$ space-time, in the original spirit of Henyey's work, but avoiding the reducibility problem.
This will be the topic of Section II. In Section III, based on the analysis developed, we spend a few words in 
an alternative interpretation of the fact that the perturbative region is insensitive to the presence of Gribov 
copies. In Section IV, we study the same problem in the presence of an instantonic background, but still in a 
3D Euclidian space-time. The analogy with the null background is straightforward, but the change in the asymptotic 
conditions imposed by the instanton alter the final conclusion. Section V is devoted to the analysis of these results, 
with an interesting possibility of absence of Gribov copies inside the horizon for gauge configurations with non 
vanishing winding number in $R^3$ space-time.

\section{Gribov Copies Inside The Horizon}

When presenting the Morse theory point of view of the Gribov's problem, we mentioned the study of the
constant copies coming from Henyey's configuration done in \cite{1992}. But, actually, the first 
example of copies in the literature which can be associated to Morse theory is the original example 
given by Gribov \cite{Gribov:1977wm,silvioreview}. This example showed that to any given field in  $C_{n-1}$ 
next to the  boundary  $l_{n}$ there is a Gribov copy in $C_{n}$ close to the same boundary. The demonstration 
of this starts from a generic boundary configuration $C$ in $l_{n}$ where a continuous first order small 
displacement is applied. Then, the same displacement is applied to the (infinitesimal) copy of $C$ in $l_{n}$, 
let us call it $\tilde{C}$. Now, demanding that the fields so obtained, $A$ and $\tilde{A}$ respectively, satisfy 
the gauge condition, one can show that they are gauge copies of each other, and that they belong to the different 
regions, internal and external to the boundary $l_{n}$. So, 
they are Gribov copies in $C_{n-1}$ and $C_{n}$ 
\cite{Gribov:1977wm,silvioreview}. The displacement used in this 
demonstration is just the kind of continuous deformation which is encompassed by the analogy with Morse theory. 
It meets a maximum and a minimum along the direction described by the gauge transformation connecting $A$ 
and $\tilde{A}$ (in the other directions in the gauge orbit they both have $n-1$ negative eigenvalues in 
order to be localized in such regions). The difference between this case, where only one copy is generated 
inside the boundary, and the example of the two constant copies of \cite{1992} is that now the deformation 
is of first order in the gauge parameter used to expand the configurations around the boundary \cite{silvioreview} 
while in \cite{1992} the first order term is null, and the expansion begins in second order \cite{9711070}. 
This vanishing of the first order is in fact a necessary condition in the deformation of a minimum configuration, 
in accordance to the Morse description. In this process we certainly cross the boundary in the way to generate 
a local maximum, such that the boundary configuration that needs to be described must still be a local minimum. 
This is only possible if the first order term vanishes. 
We will use this fact to start our analysis from the boundary configuration.

Understanding this point, the configuration that we should look for must comply with this condition, 
of cancelling this first order contribution, together with the necessary condition of avoiding 
reducibility, as we have seen. Fortunately, in the case of a $SU(2)$ gauge theory in Euclidian 
$R^3$ space-time, such example can be obtained among the generic spherically symmetric 
configurations \cite{Gribov:1977wm}
\begin{eqnarray}
A_{i}=f_{1}(r){\partial \hat n \over \partial x_i}+ f_{2}(r)\hat n{\partial \hat n 
\over \partial x_i}+f_{3}(r)\hat n n_i ,
\label{eqai}
\end{eqnarray}
with the notations
\begin{equation}
n_{i}={x_{i}\over r } ,\hspace{1cm}\hat n = in_{i} \sigma_i ,\hspace{1cm}{\hat n}^2 = -1 .
\label{91011}
\end{equation}
The $\sigma_i$ are Pauli matrices with

\begin{eqnarray}
\sigma_i\sigma_j = i\epsilon_{ijk}\sigma_k + \delta_{ij} .
\label{matrizpauli}
\end{eqnarray}
The gauge condition will be the standard
\begin{eqnarray}
\partial_i A_i = 0 ,
\label{landau}
\end{eqnarray}
which, when applied to the $A_{i}$ of (\ref{eqai}), gives  
\begin{eqnarray}
f{_3}^{'} +{{2f_3}\over r}-{{2f_1}\over r^2}= 0 .
\label{eqlandau}
\end{eqnarray}
In this last expression, and from this point on, the number of primes specifies the number of 
derivatives in relation to the argument of the function.
In the next step, we apply the gauge transformation which preserves the spherical symmetry

\begin{equation}
S=e^{i \frac{\lambda\alpha(r)}{2}\overrightarrow{n}\ldotp \overrightarrow{\sigma}} 
= cos(\frac{\lambda\alpha(r)}{2}) +i \overrightarrow{n}\ldotp \overrightarrow{\sigma}sin(\frac{\lambda\alpha(r)}{2})
\label{spherical}
\end{equation}
to the configuration (\ref{eqai}), generating the transformed field
\begin{eqnarray}
\tilde{A}_i &=& i\left(f_1\cos(\lambda\alpha)+(f_2+\frac{1}{2})\sin(\lambda\alpha)\right)\frac{\sigma_i}{r} + \nonumber \\
&+& i\left(rf_3 +r\frac{\lambda\alpha'}{2}- f_1\cos(\lambda\alpha)
- (f_2+\frac{1}{2})\sin(\lambda\alpha)\right) \frac{x_i x_j\sigma_j}{r^{3}}+ \nonumber \\
&+& i \left( (f_{2}+\frac{1}{2})\cos(\lambda\alpha) - f_{1}\sin(\lambda\alpha)-\frac{1}{2}\right) \epsilon_{ijk}\frac{x_j\sigma_k}{r^{2}}.
\label{(16)}
\end{eqnarray}
In the last expressions, $\lambda$ is a constant parameter that has been introduced to express the perturbative 
expansion away from the boundary.
If we demand that $\tilde{A}_i$ also satisfies the gauge condition (\ref{landau}), and using (\ref{eqlandau}), we get
\begin{eqnarray}
\lambda\alpha^{''}+{2\over r}\lambda\alpha^{'} -{4\over r^2}\left( (f_2+\frac{1}{2})\sin\lambda \alpha +f_1(\cos \lambda\alpha -1)\right) =0 .
\label{(17)}
\end{eqnarray}
Now, in order to meet the condition that these configurations must have vanishing first 
order contributions, and in this way generate multiple copies, the last equation must have a 
definite parity for the change $\alpha$ into $-\alpha$. This can be obtained by making $f_1$ null 
(and $f_3$ also, for simplicity, to satisfy (\ref{eqlandau})). With this choice, and renaming $f_2=f$, eq.(\ref{eqai}) can 
be rewritten as
 \begin{eqnarray}
A_i ={i \over r^2}\epsilon_{ijk}x_j\sigma_kf(r) .
\label{(18)}
\end{eqnarray}
This is finally our starting configuration. We observe that in this format, it is irreducible in $SU(2)$.
Its gauge copy becomes
\begin{eqnarray}
\tilde{A}_i &=& i(f+\frac{1}{2})\sin(\lambda\alpha)\frac{\sigma_{i}}{r} 
+ i\left( \frac{r\lambda\alpha'}{2}-(f+\frac{1}{2})\sin(\lambda\alpha)\right) \frac{x_i x_j\sigma_j}{r^{3}}+\nonumber \\
&+& i \left( (f+\frac{1}{2})\cos(\lambda\alpha)-\frac{1}{2}\right) \epsilon_{ijk}\frac{x_j\sigma_k}{r^{2}},
\label{gaugecopy}
\end{eqnarray}
%\begin{eqnarray}
%\tilde{A}_i &=& i(f+\frac{1}{2})sin(\lambda\alpha){\frac{\sigma_i}{r}} + i\left( {{r{\lambda\alpha^'}}\over 2} - (f +\frac{1}{2})sin(\lambda %\alpha)\right) {x_ix_j\sigma_j\over r^3}+ \nonumber \\
%&+&i\left((f +{1\over 2})cos(\lambda\alpha) -{1\over 2}\right) \epsilon_{ijk}{{x_j\sigma_k}\over r^2} ,
%\label{gaugecopy}
and eq. (\ref{(17)}) becomes the pendulum equation
\begin{eqnarray}
\lambda\alpha^{''}(r)+{2\over r}\lambda\alpha^{'}(r) -{4\over r^2}(f+{1\over 2})\sin\lambda \alpha=0 .
\label{20}
\end{eqnarray}
Actually, with a different motivation, such configurations were also considered in 
\cite{Gribov:1977wm,silvioreview}. 
The point is that, having assured the definite parity by the change $\alpha$ into $-\alpha$ in the last
pendulum equation, we see that both situations will be solutions to this equation. This does not change 
the initial configuration (\ref{(18)}), but we will get in the end two different copies (\ref{gaugecopy}), depending on the 
choice $\alpha$ or $-\alpha$. Also, the pendulum equation so obtained, eq (\ref{20}), allows us to rewrite 
it in a form analogous to the procedure used by Henyey \cite{henyey}
\begin{eqnarray}
f(r)={r^2\over {4\sin\lambda \alpha}}\left(  \lambda\alpha^{''}+{2\over r}\lambda\alpha^{'}\right) -{1\over 2} .
\label{21}
\end{eqnarray}
At this point we need to stress that once we intend to maintain ourselves in the same connected region of the 
gauge orbit of (\ref{(18)}), the gauge transformation (\ref{spherical}) should carry a vanishing winding number. 
This characterizes the small gauge transformations \cite{ rajaraman}, which implies that $\alpha(r)$ is regular 
in all points and 
\begin{eqnarray}
\alpha (r\rightarrow \infty) \rightarrow 0 .
\label{22}
\end{eqnarray}
Such transformations are continuously deformable into the identity, and equivalence according to them is 
which actually defines the homotopic classification of the gauge configurations. On the other hand, gauge fields 
belonging to different topological sectors can be related by large gauge transformations, carrying non vanishing 
winding number. As we mentioned in the introduction, originally the presence of copies inside the horizon was 
exclusively associated to these large transformations \cite{sciuto,semenov:1982,1992}. 
Then, when we say that our intention is to show the presence of copies inside the horizon with a null winding 
number, we are in fact stating that these copies are related by a small gauge transformation.

Now, equation (\ref{21}) can be taken as the basis for the analogy with Morse theory developed 
in \cite{1992}. As we explained in the case of the constant gauge copies of \cite{1992}, or in the 
case of the copies next to any Gribov boundary \cite{Gribov:1977wm}, the process of deformation always passes 
by copies on a boundary $l_{n}$. These copies are related by an infinitesimal gauge transformation $S^0$ . 
Then, we first take this infinitesimal limit in our case of the gauge transformation (\ref{spherical}) and write
\begin{eqnarray}
S^0 = 1+i{\lambda\alpha \over 2}\overrightarrow{n}\cdot\overrightarrow{\sigma} .
\label{szero}
\end{eqnarray}
In the same limit, eq. (\ref{21}) becomes
\begin{eqnarray}
f^0={r^2\over {4\alpha}}\left(  \alpha^{''}+{2\over r}\alpha^{'}\right) -{1\over 2} .
\label{f0}
\end{eqnarray}
From (\ref{gaugecopy}), the infinitesimal transformation (\ref{szero}) generates the boundary copy
\begin{eqnarray}
\tilde{A^0}_{i} = {A^0}_{i}+{D^0}_{i}(i{\lambda\alpha \over 2}\overrightarrow{n}\cdot\overrightarrow{\sigma})
={A^0}_{i}+\lambda{D^0}_{i}\omega ,
\label{atildezero}
\end{eqnarray}
where obviously
\begin{eqnarray}
{A^0}_{i} = {i\over r^2}\epsilon_{ijk}x_j\sigma_kf^0(r) ,\hspace{1cm} \omega 
= i{\alpha \over 2}\overrightarrow{n}\cdot\overrightarrow{\sigma} ,
\label{26}
\end{eqnarray}
and
\begin{eqnarray}
{D^0}_i=\partial_i +[{A^0}_i, ] .
\label{devcov}
\end{eqnarray}
Using these expressions, one can easily show that
\begin{eqnarray}
-\partial_i{D^0}_i\omega=0 .
\label{28}
\end{eqnarray}
This confirms that ${A^0}_i$ is in a Gribov boundary, with $\omega$ as the eigenvector with null eigenvalue 
for the Faddeev-Popov operator, and $\tilde{A^0}_{i}$ is its copy on the boundary. We just need to remember 
that $\omega$ must be normalizable in order to be a legitimate zero mode of the Faddeev-Popov operator. 
This condition implies 
\begin{eqnarray}
\lim_{r\rightarrow \infty}r^3 \alpha^{2} (r) \rightarrow 0 ,
\label{29}
\end{eqnarray}
which is consistent, and more restrictive, than (\ref{22}).
The next step is to calculate the eigenvalue $\epsilon$ for our initial field ${A}_i$ 
\begin{eqnarray}
-\partial ^2 \psi-\partial_i [A_i,\psi] =\epsilon\psi .
\label{30}
\end{eqnarray}
We first expand ${A}_i$ as a perturbative series  in $\lambda$ from the boundary solution ${A^0}_i$, 
using (\ref{(18)}), (21) and (\ref{26}),
 
\begin{eqnarray}
A_i &=& {A^0}_i+\lambda^2\left( i\epsilon_{ijk}x_j\sigma_k{\alpha\over 4!}(\alpha^{''}+{2\over r}\alpha^{'})\right) +o(\lambda^4) \nonumber \\
&=&{A^0}_i+\lambda^2a_i +o(\lambda^4) ,
\label{31}
\end{eqnarray}
which obviously satisfies the gauge condition. We can compute the eigenvalue in (\ref{30}) 
by making an analogy with perturbation theory in quantum mechanics \cite{silvioreview}. 
We substitute the expansion of (\ref{31}) inside (\ref{30}) to write

\begin{equation}
-\partial^{2}\psi -\partial_{i}[{A^{0}}_{i}+\lambda^{2}a_{i},\psi ]=\epsilon(a)\psi.
\label{32}
\end{equation}
This equation can be interpreted as a time independent Schroedinger equation with a small perturbation 
in relation to the equation with zero energy (\ref{28}). The energy $\epsilon(a)$ can be evaluated by the 
expectation value of the perturbative element in the non perturbed state

\begin{equation}
\epsilon(a) =-{Tr\int d^3x\omega^\dagger\lambda^2\partial_i[a_i,\omega]\over {Tr\int d^3x\omega^\dagger \omega}} .
\label{33}
\end{equation}
Getting $a_i$ from (\ref{31}) and $\omega$ from (\ref{26}), we can show that
\begin{eqnarray}
\epsilon(a) ={\lambda^2\over {6\int drr^2\alpha^2}}\left\lbrace [r^2\alpha^3\alpha^{'}]_{0}^\infty
-3\int drr^2\alpha^2(\alpha^{'})^2\right\rbrace .
\label{34}
\end{eqnarray}
We can now use the regularity condition (\ref{29}) to establish 
the vanishing of the first term in the asymptotic limits, and as the second one is strictly negative, we get 
\begin{equation}
\epsilon(a) <0 .
\label{35}
\end{equation}
This informs us that the eigenvalue of the Faddev-Popov operator for our starting gauge configuration (\ref{31}) is negative, 
which implies that this configuration is outside the boundary $l_n$ where  ${A^0}_i$ of (\ref{26}) is located.
We need now to calculate the eigenvalue for the gauge copies generated by the transformation (\ref{spherical}) with the expansion up to second order in $\lambda$
\begin{eqnarray}
S=1+\lambda \omega+ \lambda^2 (-{\alpha ^2\over 8})+o(\lambda^3).
\label{36}
\end{eqnarray}
This expansion can be applied to eq. (\ref{gaugecopy}), and using (\ref{atildezero}), we obtain
\begin{eqnarray}
\tilde{A}_{i} = \tilde {A^0}_{i}-2\lambda^2a_i+o(\lambda^4) .
\label{37}
\end{eqnarray}
These copies $\tilde{A}_{i}(\pm\alpha)$ will satisfy the eigenvalue equation,
\begin{eqnarray}
-\partial ^2 \tilde\psi-\partial_i [\tilde{A^0}_i-2\lambda^2a_i,\tilde\psi] =\epsilon(\pm\alpha)\tilde\psi ,
\label{38}
\end{eqnarray}
and, as $\tilde{A^0}$ is a boundary configuration with the same zero mode $\omega$
\begin{eqnarray}
-\partial ^2 \omega -\partial_i[\tilde{A^0}_i,\omega]=0 ,
\label{39}
\end{eqnarray}
we immediately arrive at
\begin{eqnarray}
\epsilon(\pm\alpha)=-2\epsilon(a) .
\label{40}
\end{eqnarray}
The conclusion is that the small copies  $\tilde{A}_{i}(\alpha)$ and $\tilde{A}_{i}(-\alpha)$ are located inside the Gribov boundary  $l_n$.

This does not finish our work here. Evidently, we still need to prove that we can find a boundary 
configuration ${A^0}_i$ of (\ref{26}) in the horizon $l_1$. This means that for this configuration all other 
Faddev-Popov eigenvalues must be positive, and that the regularity asymptotic conditions must be satisfied. 
Let us start by rewriting the eigenvalue equation for the Faddev-Popov operator 
\begin{eqnarray}
-\partial_i {D^0}_i\cdot=-\partial^2\cdot-[{i\over r^2}\epsilon_{ijk}x_j\sigma_kf^0(r),\partial_i\cdot] 
\label{41}
\end{eqnarray}
acting on a generic eigenvector $\omega_a\sigma_a$, in the form
\begin{eqnarray}
-\partial ^2 \omega_a+{2f^0\over r^2}\hat L_{ab}\omega_b=\epsilon\omega_a ,
\label{42}
\end{eqnarray}
where $\hat L_{ab}$ can be read as the angular momentum operator
\begin{eqnarray}
\hat L_{ab}=x_a\partial_b - x_b\partial_a .
\label{43}
\end{eqnarray}
In $3D$ we also have that
\begin{eqnarray}
\partial ^2 = {1\over r^2}\partial_r(r^2\partial_r)+{{\hat L}^2\over r^2} .
\label{44}
\end{eqnarray}
We now use some results of \cite{nosso}, where the following properties of the basis $Q_{b_1....b_l}$ were shown
\begin{eqnarray}
\partial_r Q_{b_1....b_l}&=&0 , \nonumber \\ 
\hat{L}^{2}Q_{b_1....b_l}&=&-l(l+1)Q_{b_1....b_l} ,\nonumber \\
\hat{L}_{ab} Q_{c_1....c_l}&=&-\sum^{l}_{i=1}\delta_{c_{i[a}}Q_{c_1...c_{i-1}b]c_{i+1}...c_l}.
\label{454647}
\end{eqnarray}
In this last expression, the anti-symmetrization involves only the indices $a$ and $b$. These results allows us to expand the eigenvectors $\omega_a$ in such basis
\begin{eqnarray}
\omega_a =\tau_{ab_1...b_l}Q_{b_1...b_l} ,
\label{48}
\end{eqnarray}
where $\tau=\tau(r)$, and then
\begin{eqnarray}
{\hat L}^2\tau = 0,
\label{49}
\end{eqnarray}
and
\begin{eqnarray}
{\hat L}_{ab}\tau =0 .
\label{50}
\end{eqnarray}

The Faddeev-Popov condition (\ref{42}) becomes
\begin{eqnarray}
&&\left( -\partial_r^2 - \frac{2}{r}\partial_r + {l(l+1)\over r^2}\right) \tau_{ab_1...b_l}Q_{b_1...b_l} \nonumber \\
&+&2 \frac{f^{0}}{{r^2}}\left( \sum_i\tau_{bb_1...b_{i-1}bb_{i+1}..b_l}Q_{b_1...b_{i-1}ab_{i+1}....b_l}\right) \nonumber \\ 
&-&2\frac{f^{0}}{{r^2}}\left( \sum_l\tau_{bb_1...b_{i-1}ab_{i+1}..b_l}Q_{b_1...b_{i-1}bb_{i+1}....b_l}\right)  \nonumber \\
&=&\epsilon\tau_{ab_1...b_l}Q_{b_1..b_l} .
\label{51}
\end{eqnarray}

If we take the case $l=1$, where \cite{nosso}
\begin{eqnarray}
Q_{b_1}={x_{b_1}\over r},
\label{52}
\end{eqnarray}
and if in particular we make
\begin{eqnarray}
\tau_{ab_1}=\delta_{ab_1}\alpha(r),
\label{53}
\end{eqnarray}
we see that equation (\ref{51}) simplifies to
\begin{eqnarray}
\left( -\partial^2_r-{2\over r}\partial_r+{{2+4f^0}\over r^2}\right) \alpha=\epsilon\alpha .
\label{54}
\end{eqnarray}
It is immediate to confirm that in this case $\epsilon=0$, by taking $f^0$ given in (\ref{f0}). This happens for 
any normalizable $\alpha(r)$ because, for the $l=1$ case, we are describing the zero mode $\omega$ of (\ref{41}), 
as can be seen by substituting (\ref{52}) and (\ref{53}) into (\ref{48}), and comparing with (\ref{28}).

The question that we have to answer is if there exists a $f^0$ such that only positive 
eigenvalues, $\epsilon>0$, are admissible in (\ref{42}), for any $l$, beyond the zero mode just described. 
This would prove that the corresponding boundary configuration ${A^0}_i$ would be in fact an horizon configuration. 
In our $SU(2)$ study, eq. (\ref{51}) is greatly simplified, as we do not have symmetric tensors in the group in order 
to build a $\tau_{ab_1...b_l}$ with more than 2 indices. The fact that the tensors $Q_{b_1....b_l}$ are 
traceless \cite{nosso} is also a limitation which leads to the conclusion that, for $SU(2)$, only the $l=1$ case 
is allowed. Then, in (\ref{51}), we use (\ref{52}), but now with a general 
\begin{eqnarray}
\tau_{ab_1}=\delta_{ab_1}\tau(r).
\label{55}
\end{eqnarray}
We get
\begin{eqnarray}
\left( -\partial_r^2 -{2\over r}\partial_r + {{2+4f^0}\over r^2}\right) \tau=\epsilon\tau .
\label{56}
\end{eqnarray}

Once more we can study (\ref{56})  by the analogy with a radial Schroedinger equation. 
Its potential will be defined by the choice of the function $\alpha(r)$ in $f^0$ of (\ref{f0}). 
If we can find a $\alpha(r)$ without nodes (with the possible exception of the origin and in the infinity), 
we will assure that the solution $\tau=\alpha$ with $\epsilon=0$ of (\ref{54}) is the solution with lower energy 
for the potential well that is formed. This was shown for the specific case of $l=1$, but as we have just argued, 
this is the only possible case for $SU(2)$. Then, if we show a $\alpha(r)$ without nodes and that satisfies 
the asymptotic conditions, we will assure that any other possible solution $\tau\neq\alpha$ will have  $\epsilon>0$ in (\ref{56}).

Let us extract the conditions on $\alpha(r)$. Together with the restrictions already expressed in (\ref{22}) and (\ref{29}), 
we have also those coming from the imposition that the configurations (\ref{atildezero}) and (\ref{26}), defined after $\alpha(r)$, 
should be regular. This condition is not indispensable, as singular field configurations can still lead to finite 
actions (see \cite{nosso} for explicit examples). Anyway, asymptotically, we will adopt an even stronger restriction over 
the fields ${A^0}_i$ and $\tilde {A^0}_{i}$,
\begin{eqnarray}
\lim_{r\rightarrow \infty}(rA^0_i)&=&0  ,\nonumber \\
\lim_{r\rightarrow \infty}(r{\tilde A}^0_i)&=&0 .
\label{57}
\end{eqnarray}
This condition appears in the literature named as strong boundary condition (SBC) \cite{sciuto}. Together with the regularity of these configurations on the origin, they imply
\begin{eqnarray}
\lim_{r \rightarrow 0}f^0&=&o(r) ,\nonumber \\
\lim_{r \rightarrow \infty}f^0&=&0 .
\label{5859}
\end{eqnarray}
We also write here the conditions coming from the fact that the zero mode $\omega$ must be normalizable 
(one of them is eq. (\ref{29}))
\begin{eqnarray}
\lim_{r \rightarrow 0}(r^3\alpha^2)&=&0 ,\nonumber \\
\lim_{r \rightarrow \infty}(r^3\alpha^2)&=&0 .
\label{6061}
\end{eqnarray}
If we start substituting the ansatz for an $\alpha(r)$ without nodes
\begin{eqnarray}
{\alpha(r)} = {{kr^m}\over {(r^2+{r_0}^2)}^n}
\label{62}
\end{eqnarray}
into equation (\ref{f0}), we see that (\ref{5859}) makes $m=1$ or  $m=-2$, but condition (\ref{6061}) eliminates this second option. Finally, (\ref{5859}) implies $n={3\over 2}$, which also satisfies (\ref{6061}). We find in the end 
\begin{eqnarray}
\alpha(r)={{kr}\over {(r^2+{r_0}^2)^{3/2}}} ,
\label{63}
\end{eqnarray}
and
\begin{eqnarray}
f^0(r)=-{15\over 4} {{r_0^2r^2}\over {(r^2+{r_0}^2)^2}} .
\label{64}
\end{eqnarray}
 
The conclusion is that the ${A^0}_i$ of (\ref{26}) given by this $f^0(r)$, 
 \begin{eqnarray}
{A^0}_{i} = -{15 i\over 4} {{r_0^2}\over {(r^2+{r_0}^2)^2}}\epsilon_{ijk}x_j\sigma_k ,
\label{65}
\end{eqnarray}
is an horizon configuration. As a last observation, in order to avoid a singularity due to the presence of the $sin\lambda\alpha$ in the denominator of (\ref{21}), we may impose
\begin{eqnarray}
\lambda\alpha < \pi \hspace{1cm} \Longrightarrow \hspace{1cm} \lambda k<{{3^{3/2}\pi r_0^2}\over 2} .
\label{66}
\end{eqnarray}

With this final form, the function $\alpha(r)$ resembles that described by Henyey \cite{henyey}. But now, the initial configuration (\ref{(18)}) is not reducible. 
Consequently, the two gauge copies $\tilde{A}_i(\alpha)$ and  $\tilde{A}_i(-\alpha)$ from equation (\ref{gaugecopy}) are legitimate Gribov copies inside the horizon.

\section{Brief Commentary On The Perturbative Region}
The question of why the QCD perturbative region is insensitive to the Gribov copies has received several different 
answers since the presentation of Gribov's problem. We mentioned in the Introduction, for example, the belief that such copies would always be associated to large gauge transformations, and in this way effects of the corrections demanded by this problem would not affect the perturbative calculations. As we know, this explanation does not stand anymore.

We can also find the argument that the zero modes of the Faddev-Popov operator do not couple to the physical 
spectrum \cite{kaku}. This can be accepted as part of the explanation, but the fact is that Gribov copies are 
not restricted to infinitesimal boundary copies, which are associated to the zero modes.  The existence of the 
finite copies shown by Henyey \cite{henyey}, not included in this subspace, show that this argument is incomplete.  

We may cite the point of view that the corrections coming from the implementation of Gribov's ideas in the action 
functional display the property of becoming negligible in the UV part of the spectrum, where we expect the perturbative 
approximation to hold, by the asymptotic freedom of QCD's gauge coupling (details on the field theory implementing Gribov 
can be found in \cite{silvio,silvio2,silvio3,silvio4} and references therein). Certainly this is an important feature of such theory, but this is 
a conclusion obtained  \textit{a posteriori}. It can be seen as a guidance along its construction, rather than an 
inevitable effect.

There is then an improvement of the first argument, based on the fact that the gauge copies of the perturbative vacuum 
belong to different topological sectors \cite{amati}. Then they cannot be accessed perturbatively. But, again, the 
existence of Gribov copies with vanishing winding number,\cite{henyey}, compromise the functional integrals around 
this vacuum (another criticism can be found in \cite{astorino}, where some examples of trivial copies of the vacuum 
are built for curved spaces, see also \cite{canfora} ). However, a further development, showing that the wave functionals are localized 
around $A=0$ for weak coupling, and that their spread in configuration space is proportional to the gauge 
coupling \cite{luscher}, gave new support to this point of view.

Our intention here is just to give an alternative point of view, and in a certain sense, glue together these results. 
In the process of calculating the copies inside the boundary, we made use of a perturbative expansion around the zero mode 
configurations, eqs. (\ref{31}) and (\ref{37}). In fact, we just followed Gribov's original example of copies around the 
horizon, where the concept of a perturbative expansion is essential. In all these cases, an expansion parameter needs 
to be explicitly introduced to make this possible. And certainly we have such parameter already available in QCD: it 
is the gauge coupling itself. In \cite{ilderton} gauge transformations with this form were used in a context very 
similar to ours, to study spherical Gribov copies (although some conventions used differ from ours). There, the existence 
of these copies was interestingly employed to give a possible explanation for the confinement of physical colour charges 
predefined in a non perturbative way. If we stick to this idea, we understand that the scope of the theory to see different 
configurations as gauge copies is associated to the gauge coupling level. For an extremely low value of the coupling, the 
gauge freedom would be restricted to infinitesimal gauge transformations. Only boundary configurations, related by 
infinitesimal transformations, would be understood as gauge copies among those satisfying the gauge fixing condition. 
As the coupling increases, the theory begins to correlate more distant gauge configurations, reaching then the copies 
around the boundary, which are related by gauge transformations linear in the expansion parameter. One step further, the 
multiple copies of the kind we have described, of second order as can be seen in (\ref{31}) and (\ref{37}), are reached. This 
interpretation is in accordance with the result described in \cite{luscher}, and the general view exposed in the last 
paragraph. At the extreme perturbative level, one should only worry with the zero mode configurations of the boundaries. 
Then, the argument of \cite{kaku} fits nicely, showing that at this level of coupling the perturbative treatment will be 
precise, without the need to any restriction prescribed by Gribov to the configuration space. With the asymptotic freedom 
of QCD's coupling, we know that this happens for the UV limit. The imprecision will only appear as the energy drops, when 
the copies will gain relevance in any calculation.

Then, the point we want to reach is that, following this line of reasoning,
the restriction to the Gribov horizon is sufficient to characterize an intermediate energy level in the way to the deep IR. This restriction gets rid of the copies linear in the gauge coupling of the kind reported originally by Gribov and at the same time already describes a new behaviour of the theory in the IR. 
This is the range where the results coming from \cite{silvio,silvio2,silvio3,silvio4} will be relevant. But as the energy 
decreases even more, the restriction to a fundamental modular region \cite{1992}, free from the copies depending exclusively on higher orders of the coupling, probably 
becomes imperative. Actually, this is also the current vision coming from the lattice \cite{bornyakov}, where new effects in a deeper IR scale are arising.

\section{Gribov Copies In The Presence Of An Instanton}

In this section we will repeat the same exercise now in the presence of a non trivial background. We still remain in a 
3D Euclidian space. Our intention is to describe the behaviour of the horizon in a region of the configuration space with 
non vanishing Pontryagin number.

In \cite{baulieu}, it was found that gauge orbits in non trivial topological sectors contribute with a different 
multiplicity factor due to Gribov copies in relation to what happens in the trivial sector. This is also associated to 
the fact that large gauge copies can be located in different Gribov regions, which is already known for a long time in 
the case of the perturbative vacuum \cite{jackiw1978}. This implies that conclusions derived for a trivial topological 
sector cannot be naively extrapolated to a non trivial one. 

The study of Gribov copies in non trivial sectors was also the subject of \cite{bruckmann}, where zero modes of a 
single $SU(2)$ 4D instanton in a maximally abelian gauge were found. In this work, this horizon configuration was 
determined with the use of the norm functional (\ref{Adois}) adapted to this gauge fixing, but the study did not 
concluded if such configuration would bifurcate and allow for the presence of Gribov copies inside the horizon.

This environment also allowed the construction of sphalerons in the superposition of the  Gribov horizon with that 
of the fundamental modular region. The conclusion was that such configurations gathered the conditions to generate 
the kind of bifurcation predicted by the Morse approach to Gribov's problem \cite{9310005, 9511119, 9711070}. But, 
these three dimensional sphalerons are associated to the non trivial mappings among three spheres, $\pi_{3}(S^{3})$. 
This demands a non trivial topology for the three dimensional space, which must be that of a $S^{3}$. 

We will focus on different configurations, also based on a $SU(2)$ gauge group, but of the kind of the spherical ones 
described by (\ref{eqai}) for a $R^3$ Euclidian space. Non trivial configurations cannot be implemented in this case 
for a pure Yang-Mills theory. This is achieved only in the presence of a scalar field, minimally coupled to 
the $SU(2)$ potential. This is the Polyakov instanton, first described in \cite{polyakov}.

We will derive all the treatment of Section II supposing now the presence of a non trivial classical background. 
In the following we then specialize to the Polyakov configuration. We will see that the background itself will not 
change the view of the problem, but there will be implications originated from the non trivial asymptotic conditions. 
Exploring such conditions directly could be a more immediate route to the final consequences, but the background method 
is more instructive to the general development. This way, to fix our notations, we define a connection $\hat{A}_{i}$   

\begin{eqnarray}
\hat{A}_{i}=A^{cl}_{i} + A_{i} ,
\label{67}
\end{eqnarray}
which is composed of a quantum excitation $A_{i}$ on a classical background $A^{cl}_{i}$. The corresponding covariant 
derivative is
\begin{eqnarray}
\hat{D}_{i}= \partial_i + [A^{cl}_{i}, ] +[ A_{i},],
\label{68}
\end{eqnarray}
which can be interpreted as a classical covariant derivative plus the quantum covariant term
\begin{eqnarray}
\hat{D}_{i}= {D}_i + [ A_{i},] ,
\label{69}
\end{eqnarray}
where
\begin{eqnarray}
D_i= \partial_i + [A^{cl}_{i}, ] .
\label{70}
\end{eqnarray}

As anticipated, we use as the classical background Polyakov's configuration \cite{polyakov}

\begin{eqnarray}
A^{cl}_{i}= -{i\over 2}a(r)\epsilon_{ijk}{{x_j\sigma_k}\over r^2} ,
\label{71}
\end{eqnarray}
which also prescribes the scalar field
\begin{eqnarray}
\phi^{cl}= -{i\over 2}U(r){{x_a\sigma_a}\over r} , 
\label{72}
\end{eqnarray}
where the functions $a(r)$ and $U(r)$ attain the limits
\begin{eqnarray}
\lim_{r \rightarrow \infty}a(r)=1,
\label{73}
\end{eqnarray}
and
\begin{eqnarray}
\lim_{r \rightarrow \infty}U(r)=F ,
\label{74}
\end{eqnarray}
with $F$ a constant representing the minimum of the Higgs potential (see \cite{rajaraman} for details).

We then adopt the gauge fixing of Polyakov \cite{polyakov}
\begin{eqnarray}
D_{i}\hat{A}_i+[\phi^{cl},{\varphi}]=0 ,
\label{75}
\end{eqnarray}
where the field ${\varphi}$ is again composed of a classical and a quantum part
\begin{eqnarray}
{\varphi} = \phi^{cl} + \phi .
\label{76}
\end{eqnarray}

The quantum fields will follow the spherical configurations, which for the gauge field in (\ref{eqai}) may alternatively 
be described as
\begin{eqnarray}
A_i=if_1{\sigma_i\over r}+i(rf_3-f_1){{x_ix_j\sigma_j}\over r^3}+if_2\epsilon_{ijk}x_j{\sigma_k\over r^2} ,
\label{77}
\end{eqnarray}
and for the scalar field we write
\begin{eqnarray}
\phi=-{i\over 2}x_i\sigma_ih(r) .
\label{78}
\end{eqnarray}

Using (\ref{72}) and (\ref{78}), we immediately see that the scalar contribution to eq. (\ref{75}) vanishes identically
\begin{eqnarray}
[\phi^{cl},\phi]=0 . 
\label{79}
\end{eqnarray}
Then using the definitions (\ref{67}), (\ref{71}) and (\ref{77}), we get the total field
\begin{eqnarray}
\hat{A}_i=if_1{\sigma_i\over r}+i(rf_3-f_1){{x_ix_j\sigma_j}\over r^3}+i(f_2-{a\over 2})\epsilon_{ijk}x_j{\sigma_k\over r^2} .
\label{80}
\end{eqnarray}
When substituted in (\ref{75}), we see that Polyakov's gauge fixing condition is satisfied if
\begin{eqnarray}
f{_3}^{'} +{{2f_3}\over r}+{{2f_1}\over r^2}(a(r)-1)= 0 ,
\label{81}
\end{eqnarray}
which is the generalization of (\ref{eqlandau}) in the presence of the instanton.

We can search now for the gauge copies of the configuration (\ref{80}) satisfying (\ref{75}). 
We will use the same small gauge transformation of (\ref{spherical}) preserving the spherical symmetry. 
In first place, we show that the scalar sector do not contribute again, as $\varphi$ of (\ref{76}), using (\ref{72}) and (\ref{78}), 
is reductible
\begin{eqnarray}
\tilde\varphi=S^\dagger\varphi S=\varphi ,
\label{82}
\end{eqnarray}
which implies
\begin{eqnarray}
[\phi^{cl},\tilde\varphi]=0 .
\label{83}
\end{eqnarray}
The transformation of the gauge field is also straightforward, just observing that its classical part does not 
get transformed, as the true gauge transformation only acts on the quantum field indeed \cite{weinberg}
\begin{eqnarray}
\tilde{\hat{A}}_i=A_{i}^{cl}+\tilde{A}_i= S^\dagger A_iS+S^\dagger A_i^{cl}S+S^\dagger\partial_i S ,
\label{84}
\end{eqnarray}
in such a way that in the infinitesimal limit we obtain
\begin{eqnarray}
\delta A_i=\hat{D}_i\omega .
\label{85}
\end{eqnarray}

The transformed field will satisfy the gauge condition if the pendulum equation in the instanton background is satisfied
\begin{eqnarray}
\lambda\alpha^{''}+{2\over r}\lambda\alpha^{'} +{4\over r^2}(a-1)\left( f_1(cos\lambda\alpha -1)
+(f_2-{a\over 2}+{1\over 2})sin\lambda\alpha\right)  = 0,
\label{86}
\end{eqnarray}
and once more we see that taking $f_{1}=f_{3}=0$, and renaming $\hat{f}=f_2-{a\over 2}$ we obtain a pendulum equation 
associated to possible multiple copies, which can be written as 
\begin{eqnarray}
\hat{f}={{r^2}\over {4(1-a)\sin\lambda\alpha}}(\lambda\alpha^{''}+{2\over r}\lambda\alpha^{'})-{1\over 2} .
\label{87}
\end{eqnarray}

We proceed with the calculation of the eigenvalue of the Faddeev-Popov operator for these configurations, beginning with
\begin{eqnarray}
-D_i\hat{D}_i\psi-[\phi^{cl},[\varphi,\psi]]=\epsilon\psi ,
\label{88}
\end{eqnarray}
where the gauge field after the redefinitions assumes a form identical to that of the zero background (\ref{(18)})
\begin{eqnarray}
\hat{A}_{i}=i\hat{f}(r)\epsilon_{ijk}{{x_j\sigma_k}\over r^2} .
\label{89}
\end{eqnarray}
In an expansion up to second order in $\lambda$
\begin{eqnarray}
\hat{A}_{i}={\hat{A}^0}_{i}+{\lambda}^2\hat{a}_{i}+o(\lambda^4) ,
\label{90}
\end{eqnarray}
we identify the boundary configuration
\begin{eqnarray}
{\hat{A}^0}_{i}=i\hat{f}^{0}(r)\epsilon_{ijk}{{x_j\sigma_k}\over r^2} ,
\label{91}
\end{eqnarray}
with

\begin{eqnarray}
\hat{f}^{0}(r)={{r^2{ (\alpha^{''}+{2\over r}\alpha^{'}})}\over {4 \alpha}(1-a)}
-{1\over 2} ,
\label{92}
\end{eqnarray}
and the small second order displacement
\begin{eqnarray}
\hat{a}_{i}=i\epsilon_{ijk}{{x_j\sigma_k}\over 4!(1-a)}\alpha 
(\alpha^{''}+{2\over r}\alpha^{'}) .
\label{93}
\end{eqnarray}
The boundary copy associated to ${\hat{A}^0}_{i}$ is generated by  the infinitesimal expansion of the gauge 
transformation (\ref{szero}) with the same $\omega$ 
\begin{eqnarray}
\omega=i{\alpha\over 2}{x_i\sigma_i\over r} ,
\label{94}
\end{eqnarray}
which satisfies
\begin{eqnarray}
[\varphi,\omega]=0 ,
\label{95}
\end{eqnarray}
and 
\begin{eqnarray}
-D_i(\partial_{i} \omega + [{\hat{A}^0}_{i},\omega])=0 ,
\label{96}
\end{eqnarray}
showing that $\omega$ is a candidate to a zero mode, depending on its normalizability. Then, using the expansion (\ref{90}) in the eigenvalue equation (\ref{88}) up to second order, and observing (\ref{95}) and (\ref{96}), we can calculate the eigenvalue $\epsilon(\hat{a})$ as the expectation value of the perturbative correction in the non perturbed solution $\omega$
\begin{eqnarray}
\epsilon(\hat{a})=-\frac{Tr\int d^{3}x \omega^{\dagger}\lambda^{2}D_{i}[\hat{a}_{i},\omega]}{Tr\int d^{3}x \omega^{\dagger}\omega}.
\label{97}
\end{eqnarray}
Upon using (\ref{70}), (\ref{71}), (\ref{93}), and (\ref{94}), we arrive at
\begin{eqnarray}
\epsilon(\hat{a})=\frac{\lambda^2}{6}
{\int dr r^2\alpha^3(\alpha^{''}+{2\over r}\alpha^{'})\over
{\int drr^2\alpha^2}} ,
\label{98}
\end{eqnarray}
which is exactly the same expression (\ref{34}) for the Faddev-Popov eigenvalue that we obtained in the zero background. 
As $\alpha$ still characterizes a small gauge transformation (which does not change the winding number of the  gauge field), 
we similarly get 
\begin{eqnarray}
\epsilon(\hat{a}) < 0 .
\label{99}
\end{eqnarray}

The calculation of the eigenvalue for the copies $\tilde{\hat{A}}_{i}(\pm \alpha)$ of (\ref{84}) follows the same steps of the 
zero background case, eqs. (\ref{37}) to (\ref{40}), and we see that these copies are inside the boundary to which ${\hat{A}^0}_{i}$ belongs. 
Then, up to now we are led to the conclusion that the theory constructed  upon the non trivial background seems to suffer 
from the same pathologies of the trivial case.

However, beyond this point, the information that we are in a sector of the theory with a non vanishing Pontryagin index 
gains relevance. Actually, we took the option to work in a background gauge to follow the conventional formalization of 
any calculation based on instantons, but basically the fundamental information is really that coming from the asymptotic 
conditions imposed by the instanton configurations. When we get to describe an actual boundary configuration, preferably 
without nodes, as we have done for vanishing background, we must modify the SBC  which were appropriate to that case. 
Now, we want the field ${\hat{A}^0}_{i}$ in (\ref{91}) to have the asymptotic behaviour of the instanton, as it is defined in (\ref{71}) 
and (\ref{73}). We need that
\begin{eqnarray}
\lim_{r \rightarrow \infty}\hat{f}^0(r)=-{1\over 2}.
\label{100}
\end{eqnarray}
The other conditions coming from the necessary normalizability of $\alpha$,
\begin{eqnarray}
\lim_{r \rightarrow 0}r^3\alpha^2 &=&0 ,\nonumber \\
\lim_{r \rightarrow \infty}r^3\alpha^2 &=&0 ,
\label{101}
\end{eqnarray}
and from the regularity at the origin,
\begin{eqnarray}
\lim_{r \rightarrow 0}\hat{f}^0(r)&=&o(r),\nonumber \\
\lim_{r \rightarrow 0}\alpha (r)&=&o(r),
\label{102}
\end{eqnarray}
remain unchanged. Also notice that the regularity at the origin of the classical configuration (\ref{71}) demands \cite{rajaraman}
\begin{eqnarray}
\lim_{r \rightarrow 0}a(r)=0.
\label{103}
\end{eqnarray}

If we start from the same general ansatz given by (\ref{62}), the imposition of the conditions (\ref{101}) to (\ref{103}), leads us to the 
same restriction $m=1$ of the zero background case. Substituting in (\ref{92}), we get 
\begin{eqnarray}
\hat{f}^0(r)={{(n^2-{3\over 2}n)r^4-{5\over 2}nr_0^2r^2}\over {(r^2+r_0^2)^2(1-a)}}+{1\over {2(1-a)}}-{1\over 2} .
\label{104}
\end{eqnarray}
Finally, we must satisfy (\ref{100}), and remembering the instanton condition (\ref{73}), we see that only two alternatives are 
allowed: $n_+ = 1$ or $n_-={1\over 2}$. The problem is that none of the two candidates so obtained,
\begin{eqnarray}
\alpha_+={{kr}\over {(r^2+r_0^2)}}
\label{105}
\end{eqnarray}
or
\begin{eqnarray}
\alpha_-={{kr}\over {(r^2+r_0^2)^{1/2}}},
\label{106}
\end{eqnarray}
can satisfy the normalizability condition (\ref{101}). The conclusion is that the configurations obtained from the boundary 
field (\ref{91}), using (\ref{105}) or (\ref{106}), although satisfying the gauge fixing and generating the bifurcation structure predicted 
by Morse theory, are not physically admissible for lack of a normalizable zero mode.

\section{Conclusion}

In the first part of this work, we addressed the problem of actually constructing examples of Gribov's copies inside the 
horizon. We showed how the standard example of \cite{1992} cannot be considered as legitimate, as the copies become rigid 
global copies inside the horizon due to the reducibility of the starting Henyey configuration \cite{9207029}. 
We succeeded in presenting such example for a topologically trivial field, which confirms the assumption that copies 
inside the horizon would not be restricted to those coming from large gauge transformations, or topologically 
non trivial spacetimes. There is also an interesting question that could be raised here. In our present case, and also in that shown in  \cite{1992}, multiple copies are generated only \emph{inside} the horizon. We never see the possibility of a bifurcation allowing copies on the outside. One could associate this effect to the symmetries of the initial gauge configurations of these examples, and imagine that in general the bifurcation process would work in both ways from any boundary. But this is not so. In order to see this, we just need to retrace the origin of the bifurcation from the analogy with Morse theory. As reasoned in \cite{1992}, this process begins with the deformation of a critical point of the topographic functional (\ref{Adois}), such that its third order contribution in the gauge parameter expansion vanishes. Then, in this gauge orbit, an expansion of the functional (\ref{Adois}) around the boundary configuration will only start in a fourth order. If this critical configuration represents a maximum in its gauge orbit, Morse theory indicates that a bifurcation would generate copies outside the boundary, from the conservation of the Euler's characteristic (\ref{Eulercha}). But the fact is that for the $SU(2)$ group, this fourth order element is always positive definite, as proved in the appendix of \cite{zwanziger}. This corresponds to a minimum condition. Then, any bifurcation starting from an $SU(2)$ horizon configuration (with a non null contribution in its gauge orbit appearing only in the fourth order) will only possibly generate double copies inside the horizon.

The physical relevance of the existence of Gribov copies is unquestionable. And their existence inside Gribov's horizon is again a physical puzzle of undeniable relevance. Not only for the main fact that their presence can change the physical behaviour of particles described by gauge theories, but also because no one has actually any idea of how to implement a restriction on field space in order to get rid of them, and so define the Fundamental Modular Region. When we began our search, our first idea was that if we could not find these copies among spherical trivial configurations (which gather  the mathematical  conditions for not allowing global copies among them, as we explained in the text), then this could enable us to even conjecture that in the trivial sector the FMR would be all the first Gribov region, at least in a continuum  Euclidian space-time. The fact that we could not find any physical argument to avoid the construction of our example in Section II made this conjecture false. 

In the second part, after spending a few words on the repercussion of this development on the effects of Gribov's copies at the perturbative 
level, we analysed the same question of copies inside the horizon in the presence of a Polyakov instanton background. 
In the end, as we have seen, the same approach which showed the copies among trivial fields is obstructed by the special 
asymptotic behaviour demanded by this non trivial configuration. Obviously, this does not mean that such copies are absent 
in general for any non trivial sector of gauge orbits, but it induces the idea that we may have a coincidence between 
Gribov's first region and the FMR for some special conditions of the spacetime, necessary for the development of instantonic configurations. In such sectors, the confirmation of this hypothesis would allow the study of a  
gauge theory actually free of Gribov copies by implementing at the action level the restriction of the configuration 
space to the region up to the horizon \cite{silvio,silvio2,silvio3,silvio4}. And as already indicated in lattice simulations \cite{bornyakov}, we believe that this restriction to a FMR will probably unveil new phenomena in the deep infrared of Yang-Mills theory not described up to now.

\end{document}